\documentclass[submission,copyright,creativecommons]{eptcs}

\usepackage[latin1]{inputenc}
\usepackage[english]{babel}
\usepackage{wrapfig}
\usepackage{pslatex}
\usepackage{setspace}
\usepackage{float}
\usepackage{moreverb}
\usepackage{makeidx}
\usepackage[usenames]{color}
\usepackage{listings}
\usepackage{textcomp}
\usepackage{paralist}
\usepackage{xspace}
\usepackage{amsmath}
\usepackage{amsfonts}
\usepackage{stmaryrd}
\usepackage{framed}
\usepackage{rotating}
\usepackage{multirow}
\usepackage{array}
\usepackage{caption}
\usepackage{multicol}

\usepackage{listings}

\definecolor{darkgreen}{rgb}{0,0.5,0}

\RequirePackage{array}
\newenvironment{authors}[1]%
  {\begingroup
   \newcommand\estyle{}%
   \renewcommand\institute[1]%
     {\\\multicolumn{#1}{@{}c@{}}{\scriptsize\begin{tabular}[t]{@{}>{\footnotesize}c@{}}##1\end{tabular}}}%
   \renewcommand\email[1]%
     {\gdef\estyle{\footnotesize\ttfamily}\\##1\gdef\estyle{}}
   \begin{tabular}[t]{@{}*{#1}{>{\estyle}c}@{}}
  }%
  {\end{tabular}%
   \endgroup
  }

\lstdefinelanguage{its}
{morekeywords={if, =, +, >=, +, -, (, ), [, ], true, false, typedef, transition, int,GAL, abort, !, \{, \}, label, &&, ., composite, synchronization,for,gal, array},
stringstyle=\color{pink},
morecomment=[l]{//},
sensitive=false,
}
\usepackage{breakurl}             
\usepackage{underscore}           

\providecommand{\event}{MARS 2018} 

\newcommand{\code}[1] {\small\texttt{#1}\normalsize}

\DeclareGraphicsExtensions{.pdf,.jpg,.png}

\graphicspath{{Figures/}{FiguresGen/}}

\title{Modeling a Cache Coherence Protocol with the Guarded Action Language}
\author{
  \begin{authors}{3}
    Quentin L. Meunier&  Yann Thierry-Mieg & Emmanuelle Encrenaz 
\institute{Sorbonne Universit\'{e}, CNRS \\
Laboratoire d'Informatique de Paris 6\\
LIP6, F-75005 Paris, France}
\email{Quentin.Meunier@lip6.fr & Yann.Thierry-Mieg@lip6.fr &  Emmanuelle.Encrenaz@lip6.fr}
    \end{authors}
}
\def\titlerunning{Modeling a Cache Coherence Protocol with GAL}
\def\authorrunning{Q. Meunier, Y. Thierry-Mieg, E. Encrenaz}

\begin{document}
\maketitle

\begin{abstract}
We present a formal model built for verification of the hardware Tera-Scale ARchitecture (TSAR), focusing on its Distributed Hybrid Cache Coherence Protocol (DHCCP). This protocol is by nature asynchronous, concurrent and distributed, which makes classical validation of the design (e.g. through testing) difficult. We therefore applied formal methods to prove essential properties of the protocol, such as absence of deadlocks, eventual consensus, and fairness.
\end{abstract}

\section{Introduction}

Testing and simulation are unfortunately not sufficient to provide strong correctness guarantees expected from hardware designs, where patching is not an option. However, proving a design correct is a difficult process despite the improvement of verification tools, as complexity of hardware designs has also grown along with Moore's law.

The TSAR (Tera-Scale ARchitecture) shared memory architecture~\cite{tsar} studied in this paper is a general-purpose multicore architecture, in which cache-coherence is entirely supported by the hardware. The main technical challenge of this architecture is scalability, as it is intended to integrate up to 1024 cores. Its embedded cache-coherence protocol is a key architectural point, and has been designed scale well when the number of cores grows.

To formally prove properties of the designed protocol, named Distributed Hybrid Cache Coherence Protocol (DHCCP), we built several formal models over a number of student internships, designed to investigate how different model-checking tools could address our need. This case study is the result of a collaboration between the experts in systems-on-chip design that developed TSAR, and experts in formal verification that provide model-checking tools.

We first present the DHCCP protocol and the relevant characteristics of the hardware that must be captured by the semantics of the model. We then present the formal models that we built over time in Promela, Divine and GAL, together with the results we were able to obtain.

All these models are made available as part of this submission process and will be accessible in the MARS repository as well as on github\footnote{\url{https://github.com/lip6/TSAR-DHCCP}}.

\section{TSAR Hardware Architecture and Coherence Protocol}

\label{sec:hardware}
\begin{figure}[htbp]
  \begin{center}
    \includegraphics[width=0.8\columnwidth]{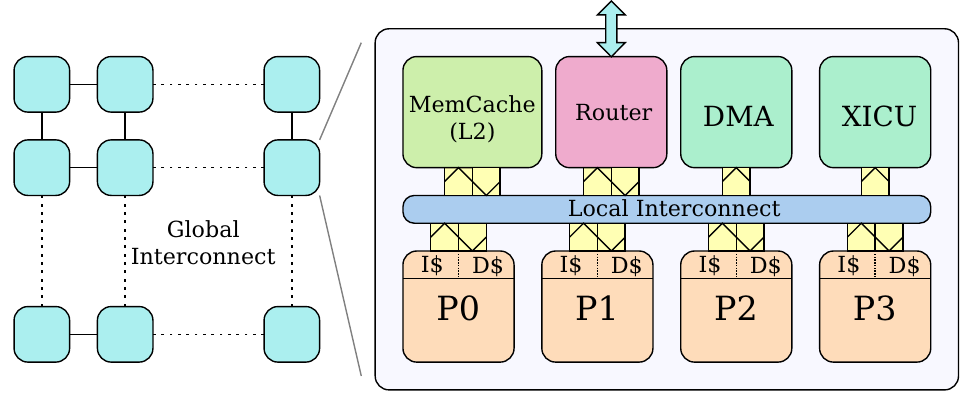}
    \caption{\textbf{Tsar Architecture Overview: Mesh Topology and Cluster Details}}
    \label{fig:tsaroverview}
  \end{center}
\end{figure}

\medskip
\textbf{Memory Layout.}
The architecture is clustered and has a 2D-mesh topology, as represented in Figure~\ref{fig:tsaroverview}. Each cluster typically contains 4 processor cores with their L1 caches, a local interconnect, one L2-cache bank, and some peripherals.

The set of all L2-cache banks form a distributed L2 cache: the physical address space is statically split into fixed-size segments, and each L2 cache is responsible for one segment, defined by the physical address MSB bits. As such, the L2 cache is logically shared and physically distributed, making the architecture NUCA (Non Uniform Cache Access): all processors can access all L2-cache banks, but the access time and power consumption depend on the distance between the processor and the cache bank.

\medskip
\textbf{L1--L2 Communication.}
\label{sec:interconnects}
For L1-L2 communication, TSAR uses a \textit{L1-L2 Interconnect} composed of two hierarchical levels: a \textit{Local Interconnect} for intra-cluster communication, and a \textit{Global Interconnect} for inter-cluster communication, implemented as a network-on-chip with a 2D-mesh topology. For a L1 cache, accessing a L2 located in another local interconnect results in a longer latency, but both levels implement a logically flat address space. The L1-L2 interconnect provides a built-in broadcast service used by the coherence protocol to efficiently broadcast invalidation messages and contains five independent networks.

\medskip
\textbf{Hardware Coherence Mechanism.}
\label{sec:dhccp}
For scalability purpose, TSAR implements a directory-based cache-coherence policy. From a conceptual point of view, the coherence protocol is supported by a global directory located in the L2 controller: this global directory stores the status of each cache line replicated in at least one L1 cache of the architecture. The policy between L1 and L2 caches is write-through, meaning that the L2 cache always contains the most recent value of a cache line, and there is no exclusive ownership state for a L1 cache. This global directory is physically distributed in the corresponding L2 banks.

The basic coherence mechanism is the following: when the L2 controller receives a \textsc{Write} request for a given cache line, it sends an \textsc{Update} or \textsc{Inval} request to all L1 caches containing a copy except the writer. The write request is acknowledged only when all \textsc{Update} or \textsc{Inval} transactions are completed.

The L2 cache is inclusive: a cache line present in at least one L1 cache must be present in the L2 cache. Thus for any evicted line, the corresponding copies must be invalidated. When a shared piece of data is modified, the DHCCP protocol uses two different strategies depending on the number of copies:
\begin{itemize}
  \item \textsc{MulticastUpdate}: when the number of copies is smaller than a certain threshold, called \textit{DHCCP threshold}, the L2-cache controller registers the locations of all the copies and sends an \textsc{Update} request to each concerned L1 cache.
  \item \textsc{BroadcastInval}: when the number of copies is larger than the \textit{DHCCP threshold} or if there is no room left to register the copies locations, the L2-cache controller registers only the number of copies without their location, and sends an \textsc{Inval} request to all L1 caches. Only the L1 caches which own a copy of the line must respond to this request, thus reducing the traffic.
\end{itemize}
The list of sharers of a given cache line is stored in the L2 directory with a per bank hardware heap shared between lines. A counter of sharers is also maintained in the directory entry. When the threshold of copies is exceeded or when the heap is full, the sharers list is freed and only the counter of copies is used. In this case, broadcast invalidates are used to maintain cache-coherence.

\medskip
\textbf{Types of Transactions.}
\label{sec:dhccpmessages}
Two types of transactions are defined for L1-L2 communication:

\textbf{Direct transactions}, containing the messages \code{read}, \code{write}, \code{ll} (Load-Linked), \code{sc} (Store Conditional), \code{cas} (Compare-and-Swap), along with their responses. These transactions use two separated networks for commands and responses. They are initiated by L1-cache controllers. For these transactions, the target can be any L2-cache controller.

\textbf{Coherence transactions}, containing the messages \code{cleanup} (eviction from L1 or invalidation acknowledgement), \code{clack} (CLeanup ACK), \code{multicast\_update}, \code{multicast\_inval}, \code{broadcast\_inval}, and \code{multi\_ack}. There are four types of coherence transactions, each requiring two or three steps.

\noindent\textbf{Coherence Transactions.}
\begin{itemize}
  \item A \textbf{\code{local cleanup}} is a two step transaction initiated by a L1-cache controller when it makes a cache line replacement, usually following a miss. The L1 cache sends a \code{cleanup} request indicating it is invalidating the cache line, and the L2-cache controller returns a \code{clack} response to acknowledge the line invalidation.
  \item A \textbf{\code{multicast update}} is a two step transaction initiated by the L2-cache controller when it receives a \code{write} request to a replicated cache line, for which the number of copies does not exceed the DHCCP threshold. It sends as many \code{multi\_updt} requests as the number of registered copies, but the writer. The expected response is a \code{multi\_ack} sent by each involved L1 cache. The L2-cache controller counts the number of responses to detect the completion of the \code{multicast update} transaction.
  \item A \textbf{\code{multicast invalidate}} transaction is a three step transaction initiated by a L2-cache controller when it makes a cache line replacement (following a miss) and the victim line has a number of copies smaller than the DHCCP threshold. It sends as many \code{multi\_inval} requests as the number of registered copies; each L1 cache returns a \code{cleanup} response, and the L2 cache acknowledges the invalidation with a \code{clack}. The L2-cache controller counts the number of responses to detect the completion of the \code{multicast invalidate} transaction.
  \item A \textbf{\code{broadcast invalidate}} is a three step transaction initiated by a L2-cache controller when it either replaces a line, or receives a \code{Write} request to a replicated cache line, and this cache line has a number of copies larger than the DHCCP threshold. The L2 cache sends a single message \code{broadcast\_inval}, which is dynamically replicated by the network to all L1 caches. L1 caches which have a copy must respond by a \code{cleanup} message. All these \code{cleanup} responses are counted by the L2 cache to detect the completion of the transaction. Finally, the L2-cache controller acknowledges each invalidation with a \code{clack}.
\end{itemize}

The \code{multi\_updt}, \code{multi\_inval}, \code{broadcast\_inval} and \code{clack} messages use the direction \textit{L2 $\rightarrow$ L1 Cache}, while the \code{cleanup} and \code{multi\_ack} messages use the direction \textit{L1 $\rightarrow$ L2 Cache}.

A L1-cache controller must be sure that a sent \code{cleanup} request has arrived to the L2 cache before sending a \code{read} request for the same line; otherwise, there would be a risk of inconsistency if the latter request passes the cleanup. To enforce this, L2 caches must respond to each \code{cleanup} message with a \code{clack}. In order to avoid deadlocks, \code{clack} responses must use a physically separated network. Therefore, the coherence transactions require three separated networks:
\begin{itemize}
  \item \code{cleanup} and \code{multi\_ack} share the first network.
  \item \code{multi\_updt}, \code{multi\_inval} and \code{broadcast\_inval} share the second network.
  \item \code{clack} are conveyed by the third network.
\end{itemize}

Direct transactions use two further networks: one for requests and one for responses, for a total of five independent networks.

\section{Modeled Architecture}

\textbf{Verification Objectives.}
If intensive testing is mandatory, formal verification can help detect subtle bugs due to some uncommon interleavings of messages on the different networks. In our case, one of the main challenges in the DHCCP protocol consists in counting the correct number of responses for each coherence transaction, since an incorrect count will eventually result in a deadlock.

Our goals were first to  model formally the DHCCP protocol so as to prove the absence of deadlocks. We also wanted to prove simple functional properties, such as every request receives a response, or a shared copy modification eventually leads to an invalidation or an update of the other copies. Proving these properties would greatly strengthen our confidence in the protocol.

We did not expect verification to scale up to the full 1024 core design, but that is not truly necessary due to the symmetry of DHCCP. Our main goal was then to target a platform configuration that would exhibit all the characteristics of DHCCP by working up progressively from smaller configurations. To exhibit all relevant properties we determined that we needed a threshold for DHCCP that is strictly less than the number of possible copies of a data, so all types of coherence transactions can occur, so we need to scale to a design with at least three processors. Another parameter we identified is that at least two addresses should be represented to show that no problem arises from the sharing of the channels between the different L1 caches and L2 banks.

\textbf{Architecture Parameters.}
In order to be able to verify properties, the components in the architecture need to be abstracted, and some restrictions have to be made. The modeled components are processor, L1 cache, L2 cache, memory and interconnect channels. All of these components except channels are modeled as reactive communicating automata, they have a control location or \code{state} that defines which messages they can send or receive.

As seen in the section on the interconnect, the topology is logically ``flat'' , and is modeled as such. Channels have the capability to serialize requests coming from different sources, and to route a request to different destinations.

The architecture parameters are the following:
\begin{itemize}
  \item Number of processors and their L1 caches: \texttt{NB\_PROC}.
  \item Number of cache lines in the L2 cache: \texttt{NB\_L2}. Each L2-cache bank contains only one line in the model, and thus several cache lines translate into several L2-cache banks.
  \item DHCCP threshold: \texttt{CACHE\_TH}.
\end{itemize}

This parametrized description should be preserved during the modelling to be able to easily consider configurations of increasing complexity.

There are three significant restrictions in the models we built: $1.$ the single line L2-cache bank, $2.$ the assumption that the L2 is large enough to contain all memory lines, and $3.$ the absence of data modelling. The first restriction prevents a L2-cache bank from receiving on its port requests at different addresses. This is a reasonable simplification, as there is globally not much interaction between requests targeting different cache lines in the L2 cache. The second restriction prevents several memory addresses from being in the same L2 bank; lifting this restriction  would require to modify the L2 state machine and add several variables into it (thus yielding even quicker combinatorial explosion). The third restriction means that we only model the control part of the protocol, and abstract away the content of the data in the lines of cache. Adding data is easy from a modelling point of view, and would enable verification of some coherency properties but these data values do not impact correctness of the protocol (e.g. deadlock freedom) and they would participate in state space explosion.


Table~\ref{tab:components} describes the variables we considered relevant inside each component. The processor model is very simple and sends random read or write requests at a random address. The L1 cache contains only one line, and the line validity is defined by the cache state. Each L1 cache is defined with a unique identifier. Each L2 cache contains the line address it can hold, along with arrays for the explicit list of copies, and a variable to store the copies count.
A channel interconnects two components and is modeled as a one-place buffer. There is one channel per network, i.e. 5 between L1 caches and L2 banks for the case of DHCCP.

\begin{figure}[ht!]
  \begin{center}
    \includegraphics[width=0.9\columnwidth]{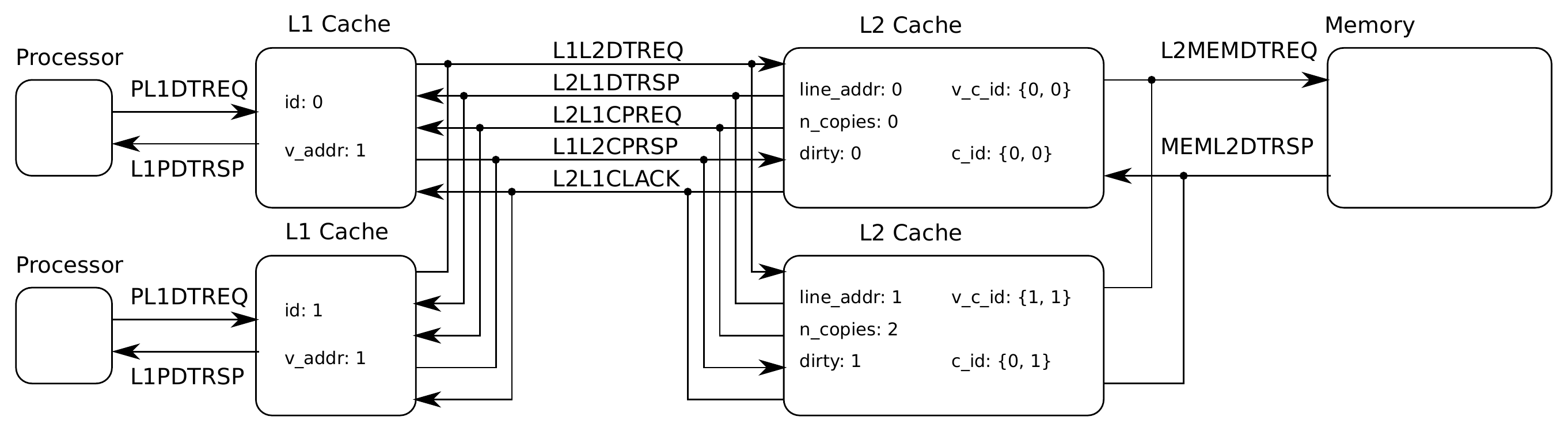}
  \end{center}
  \caption{Example of an architectural state modelisation, in which the line with address 1 has a copy in both L1 caches}
  \label{fig:typicalarch}
\end{figure}

\begin{figure}[ht!]
  \begin{center}
  \includegraphics[width=0.8\textwidth]{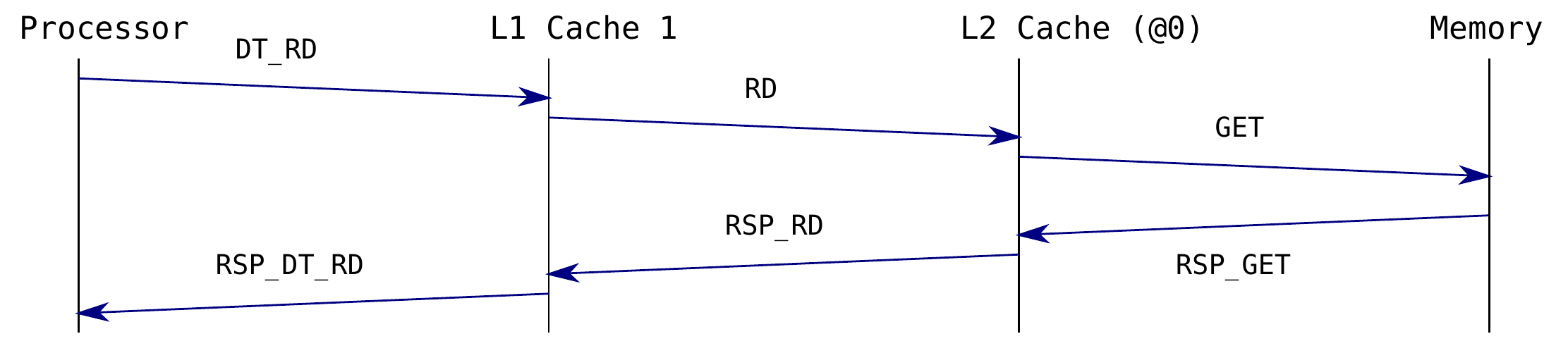}
  \captionsetup{singlelinecheck=off}
  \caption[]{Message Sequence in case of misses in the L1 and L2 caches. The messages are the following:
    \begin{enumerate}[(1)]
      \item The processor connected to the L1 cache sends a read request (message \texttt{DT\_RD}) for the line with address 0 (\texttt{PL1DTREQ} channel).
      \item The L1 cache does not own a valid copy of the line and sends a \texttt{RD} message for the line at address 0 (\texttt{L1L2DTREQ} channel).
      \item The L2 cache in charge of address 0 receives the request and, as it misses, sends a \texttt{GET} message for address 0 to the memory (\texttt{L2MEMDTREQ} channel).
      \item The memory responds with a \texttt{RSP\_GET} message to the requesting L2 (\texttt{MEML2DTRSP} channel).
      \item The L2 cache updates it state for the line and responds to the L1 with a \texttt{RSP\_RD} message (\texttt{L2L1DTRSP} channel).
      \item Upon receiving the response, the L1 cache updates its state and responds to the processor with a \texttt{RSP\_DT\_RD} message (\texttt{L1PDTRSP} channel).
    \end{enumerate}
  }
  \label{fig:examplesequence}
  \end{center}
\end{figure}

Figure~\ref{fig:typicalarch} shows a system instance with \texttt{NB\_PROC = 2}, \texttt{NB\_L2 = 2}, and \texttt{CACHE\_TH = 2}, and with some of the components variables values. Appendices~\ref{app:l1automaton} and~\ref{app:l2automaton} show the finite state machines for the L1 and L2 caches. All states are represented, but some transition actions are omitted for clarity. Appendice~\ref{app:messageschannels} describes the different messages modeled and the channels on which they are conveyed. These messages types are directly obtained from the DHCCP description in section~\ref{sec:dhccpmessages}. Figure~\ref{fig:examplesequence} shows the dynamics, through a sequence diagram describing a cache miss.

\begin{table}[hb!]
  \begin{center}
  \caption{Components modeled with their variables}
  \label{tab:components}
  \footnotesize
  \begin{tabular}{|ll|ll|}
    \hline
\multicolumn{2}{|l|}{\textbf{Processor}} & \multicolumn{2}{l|}{\textbf{Channel}} \\
    \hline
\code{state} & FSM state               & \code{address}:       & target address of the request \\
\code{addr}: & address of the last     & \code{id} \footnotesize(optional): & identifier of the request sender\\
             & request emitted         & \code{type}:          & request type \\
    \hline
\multicolumn{4}{|l|}{\textbf{L1 cache}} \\
    \hline
\multicolumn{2}{|l}{\code{state}       } & \multicolumn{2}{l|}{FSM state} \\
\multicolumn{2}{|l}{\code{id}:         } & \multicolumn{2}{l|}{identifier of the cache (different for all caches) } \\
\multicolumn{2}{|l}{\code{v\_addr}:    } & \multicolumn{2}{l|}{address contained in the cache when it is valid    } \\
\multicolumn{2}{|l}{                   } & \multicolumn{2}{l|}{(validity is encoded in the sate)                  } \\
\multicolumn{2}{|l}{\code{addr\_save}: } & \multicolumn{2}{l|}{address of the last request sent                   } \\
    \hline
\multicolumn{4}{|l|}{\textbf{L2 cache}} \\
    \hline
\multicolumn{2}{|l}{\code{state}                                       } & \multicolumn{2}{l|}{FSM state} \\
\multicolumn{2}{|l}{\code{line\_addr}:                                 } & \multicolumn{2}{l|}{address of the line mapped in this L2-cache bank} \\
\multicolumn{2}{|l}{\code{n\_copies}:                                  } & \multicolumn{2}{l|}{number of copies for this line} \\
\multicolumn{2}{|l}{\code{dirty}:                                      } & \multicolumn{2}{l|}{true if the line has been modified w.r.t. the memory} \\
\multicolumn{2}{|l}{\code{src\_save} and \code{src\_save\_clnup}:      } & \multicolumn{2}{l|}{used to save a request source id when a coherence request is needed} \\
\multicolumn{2}{|l}{\code{cpt}, \code{cpt\_clnup} and \code{rsp\_cpt}: } & \multicolumn{2}{l|}{counters used for the sending of multicast or broadcast requests} \\
\multicolumn{2}{|l}{\code{v\_c\_id[]} and \code{c\_id[]}:              } & \multicolumn{2}{l|}{arrays of size CACHE\_TH storing the explicit list of copies; } \\
\multicolumn{2}{|l}{                                                   } & \multicolumn{2}{l|}{\footnotesize if \code{v\_c\_id[i]}\footnotesize == 1, then entry \code{i} in \code{c\_id} is valid} \\
\multicolumn{2}{|l}{                                                   } & \multicolumn{2}{l|}{\footnotesize and \code{c\_id[i]} \footnotesize contains the corresponding cache identifier.} \\
    \hline
  \end{tabular}
  \end{center}
\end{table}

\section{Modeling using Communicating Process}

\medskip
\textbf{Promela Models.}
To model DHCCP we first used the Promela language and its supporting tool Spin~\cite{spin}. It offers as modeling framework asynchronous process communicating over channels. The language itself is relatively comfortable, each component is described using code-like control structures (case, loop). The tool can exhibit traces as sequence diagrams, which are particularly valuable to develop and debug the model.

We first built~\cite{najem2011} up automata of the behavior of the L1 cache and the memory controller, and abstracted the activity of processors to arbitrary read and write requests. These automata were then encoded into Promela using labels and goto.

To validate the models, we used properties encoded into ``observation'' automata, synchronized with the system. In some cases, adding these automata proved to be a problem, as they can incorrectly block the system if not well designed. Also, to observe channels we had to duplicate channels and messages (one for the true channel, one for the observer) which is quite intrusive. Overall, this observation mechanism was quite cumbersome, and participated in state space explosion.

We separately analyzed the components in simple configurations before assembling them. The simple configurations helped validate that the Promela code correctly reflected the automata. Full model-checking was however only possible on the simplest instances, with a single processor and two addresses. For the full system, we were only able to use simulation and bounded model-checking (up to roughly $10^8$ states).

This Promela model was then extended and refined with the same goals in mind~\cite{mansour2012}. We simplified and abstracted the data manipulation, and removed the observers. We were still unable to explore the full state space for three processors, reaching 270 steps in depth for $10^9$ states. We explored instead two configurations with two processors, while varying the threshold variable. We were also able to include the LL/SC (Load-link/Store-conditional) support in the TSAR architecture, leading unfortunately to much more complex automata, as more control messages were introduced.

While partial-order reduction was activated, we could not activate
$d\_step$ as every set of actions contains at least one channel
interaction. The channels themselves are shared for both writing and
reading (precluding use of $xr/xs$ keywords) since it models a bus.

In conclusion, despite some aggressive simplifications (e.g. full data abstraction), we were unable to fully verify deadlock freedom for Promela models for even the smallest truly relevant configuration, i.e. at least three processors, two addresses, and a threshold at 2 so that both multicasts and broadcasts can occur. The language and simulator were however quite comfortable to use.

\medskip
\textbf{Divine Models.}
We then built a second set of models~\cite{gharbi2013}, but this time using the Divine~\cite{divine} language.
Similarly to Promela, Divine is a language to express systems as (asynchronous) processes communicating over channels. However, it is a much simpler language with less features. Each process is described using an automaton with ``local'' variables ; transitions have a source and target state, a guard or enabling condition over variables, may send or receive messages from channels and update variables.

We chose to use Divine mostly because we wanted to try other tools than Spin. Thanks to having a relatively simple syntax and semantics, and also due to the existence of a nice set of benchmark models~\cite{beem}, several tools provided support for the language. The Divine tool coming with the language handled LTL in multi-core and distributed settings, LTSmin~\cite{ltsmin} offered support for both explicit and symbolic exploration, and building upon new results~\cite{cav13}, a prototype path to input Divine models to our symbolic model-checker ITS-Tools had been recently built.

Building the model in Divine was harder than in Promela however. On
the positive side, we have good control over the atomicity of
statements; in the Promela models, due to interaction of send/receive
actions with the \texttt{atomic} keyword, some interleavings of
independent progression of different process were still observable.
For instance, a (passive) component that consumes a message and
immediately sends an answer on a different channel could not be
modeled without the state between receive and send being materialized.

On the other hand, we fell into some difficulties to model channels; Promela let us peek at the content of a channel without consuming it, enabling a routing mechanism where an entity only consumes messages which are addressed to it. Without this mechanism, we were in fact unable to model the semantics we needed; we had to resort to explicit modeling of the channel as shared global variables, and simulating read and write operations with instructions. Divine's support for parametric modeling was also poor, since it basically recommends the use of the macro expansion tool \textit{m4}, which is not comfortable.

During this internship, we again built separate components, then progressively assembled them to build more complex configurations. We used LTL (instead of observation automata) to express expected properties of the system, under fairness constraints that force all of the processes to progress. Without fairness, most of the properties were not valid, which was expected, but unfortunately fairness is not supported by all the tools.

\begin{figure}[bthp]
  \begin{center}
    \includegraphics[width=0.5\columnwidth]{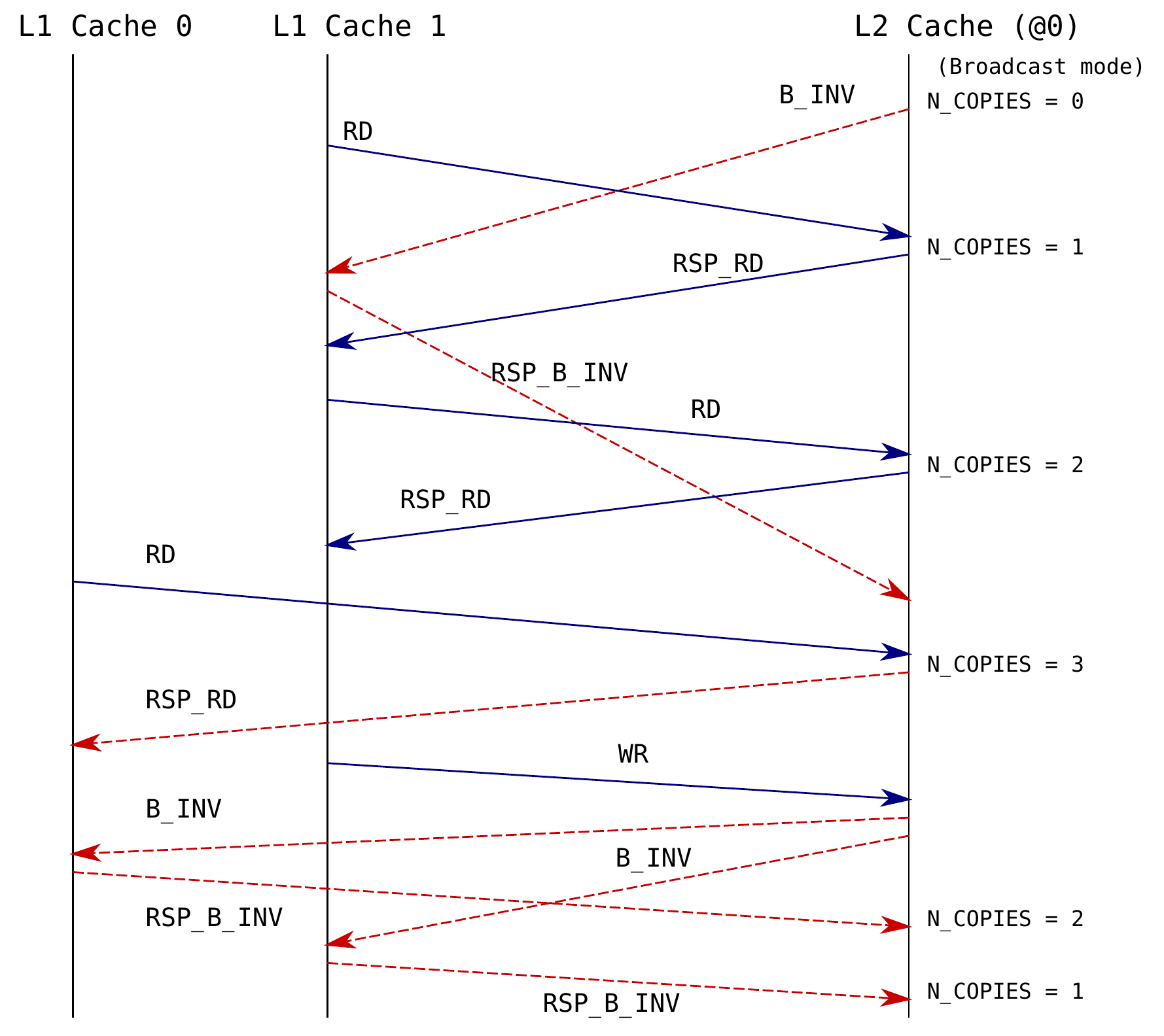}
    \caption{Part of a deadlock sequence in a previous version of the DHCCP protocol. Blue arrows represent direct messages while red arrows represent coherence messages. Messages can be received out of order due to the existence of several channels between the memory controller and the L1 cache. The \texttt{RSP\_B\_INV} message was exclusive to that version of DHCCP. At the end of the sequence, the line count in the L2 is one whereas there are no copies left.}
    \label{fig:dhccpdeadlock}
  \end{center}
\end{figure}

With these models, we were able to reproduce a real deadlock found on a previous version of the DHCCP protocol. On the configuration with two processors, two addresses and a threshold set at 1, the scenario exhibited by the sequence diagram in Figure~\ref{fig:dhccpdeadlock} could occur. It leads to an incorrect counter value, that ultimately leads to a deadlock by propagation as the head message in the FIFO channel cannot be consumed and communications lock down.

This bug was detected in the cycle-accurate TSAR prototype by its designers, using testing and simulation, but it took one year to detect this issue and correctly diagnose it. Building a solution to correct the problem was non trivial, and took another six months. In contrast, once the formal models were built, detecting it was easy, even for a relatively non expert Master 2 student, and model-checking could exhibit readable diagnostic traces. We were then able to modify our model to match the next (correct) version of the DHCCP protocol described in section~\ref{sec:dhccp}, and we checked that the deadlock issue was indeed resolved.

However, we were still unable to fully verify larger configurations with three or more processors. The use of ITS-tools was possible only for deadlock detection, as fairness constraints were unavailable to check the more complex LTL properties. The input from Divine to ITS-tools also involves several automated transformation steps, that yielded a relatively complex model due in part to channel communications modeled as shared variables, and to the loss of structural information in the transformation (i.e. from a set of processes to a single large specification). Experimentation with LTSmin was not extensive, but we measured performance similar to that of the Divine tool; again chains of transformations and relatively poor tool integration made this path less comfortable than just using Divine natively.

In conclusion, using Divine models, we were able to successfully reproduce a critical bug in the TSAR coherence protocol, and to prove that the patch did solve the problem. However, we were still unable to perform verification for larger instances with three or more processors.

\section{Modeling with GAL}

The third set of models were built using the Guarded Action Language (GAL)~\cite{gal} during a Master Thesis~\cite{zhao2015}. GAL is a language offering very fine control over the expression of concurrent semantics, with no assumptions on the existence of higher level constructs such as processes or channels and very few keywords.

\medskip
\textbf{Guarded Action Language.} GAL is a formalism supporting hierarchical descriptions of components; terminal or leaf components are specified as GAL type declarations, while composite type definitions allow to instantiate existing types (of GAL or composite nature) and synchronize these instances. A GAL specification is then composed of a set of type declarations, and a specific instance \texttt{main} which is designated as the full system. These characteristics of the language, borrowed from architectural description languages, help reuse model elements in various configurations easily.

A GAL type declaration defines a set of integer variables and fixed-size arrays of integers as variables. A state of a GAL is then a complete assignment of an integer value to each of these variables. Transitions are defined as a triplet $\langle g, l, a \rangle$, where $g$ is a boolean enabling condition or guard, $l$ is a label chosen from a finite set, and $a$ is a sequence of statements or actions that must be executed atomically. A statement can be an assignment to a variable of an expression computed over variables, or a call to a label. The call statement is resolved by finding a transition bearing the target label, whose guard is enabled, and executing it.

A composite type declaration defines a set of instances and fixed-size arrays of instances as variables. A state of a composite is then a complete assignment of a subcomponent state to each of these instances. Synchronizations are defined as a pair $\langle l, a \rangle$, where $l$ is a label chosen from a finite set, and $a$ is a sequence of call statements that must be executed atomically. A call statement has a target, which can be either the enclosing instance itself, or any nested instance. The target of the call must then evolve through a transition (or synchronization in composite case) that bears the target label.

GAL also offers parametric modeling features, in the form of parameters defined over a discrete range of values\footnote{In the syntax, parameters are distinguished from variables by a \$ sign.}. These parameters let us define parametric transitions, that correspond to a set of non parametric transitions, one per possible value of the parameters. Parameters can be used to define parametric labels, in order to model communications over discrete data types as calls to labels.

GAL is mostly designed to be the target in a model transformation process, where the specification is typically expressed using a domain specific notation such as Promela or Divine, and automatically translated to GAL for analysis purposes. Going for direct modeling in GAL however, gives us proximity to the symbolic solution engine, enabling the use of advanced features that are not built into the general purpose transformation, e.g. from Divine.

In particular, the automatic transformation loses structural information, yielding a single GAL component instead of a composition of process. The encoding of channels as global variables also prevents the transformation from automatically building a representation using synchronizations on labels.

\medskip
\textbf{Elementary Components.}
We separately developed the models of the communication channels and of the various components. For normal components, we model automata by defining a \texttt{state} variable, then adding a GAL transition for every transition of the automata.

For channels, we built a GAL with enough variables to store the message, and a boolean flag to test if the channel is full. For each possible message going through the channel (which must be a finite domain) we generate two transitions \textit{read} and \textit{write} with a distinct label for each of them. In this example (see Figure~\ref{fig:changal}) the channel messages carry a target address (among possible memory slots) and a message type (among 20 possible values, e.g. update request). The read operation also flushes the state of the channel, to prevent this unreadable state information from participating in the state space explosion.

\begin{figure}
\centering
	\begin{multicols}{2}
	\begin{lstlisting}
typedef addr_t = 0 .. $NB_L2 - 1 ;
typedef type_t = 0 .. 19 ;
typedef id_t = 0 .. $NB_PROC - 1 ;

gal ChannelAddrType {
 int isFull = 0 ;
 int addr = 0 ;
 int type = 0 ;
 transition read
  (addr_t $addr, type_t $rtype)
  [isFull == 1 && addr == $addr
   && type == $rtype]
  label "read" ($addr, $rtype) {
   isFull = 0 ;
   addr = 0 ;
   type = 0 ;
 }
 transition write
  (addr_t $addr, type_t $wtype)
  [isFull == 0]
  label "write" ($addr, $wtype) {
   isFull = 1 ;
   addr = $addr ;
   type = $wtype ;
 }
}
gal CacheL1 {
 int state = $INIT ; // state in the automaton
 int v_addr = 0 ; // address in cache if VALID
 int addr_save = 0 ; // saves the address
                     // of a sent request
 int id; // fixed identifier of this L1 cache

 transition t_init (id_t $id)
  [state == $INIT]
  label "init" ($id) {
   state = $L1_EMPTY;
   id = $id;
 }
 transition t_Empty_WriteWaitEmpty
  (id_t $id, addr_t $addr)
  [state == $L1_EMPTY && id == $id ]
  label "read_PL1DTREQ_write_L1L2DTREQ"
     ($id, $addr, $DT_WR, $WR) {
    state = $L1_WRITE_WAIT_EMPTY ;
    addr_save = $addr;
 }
...
\end{lstlisting}
 	\end{multicols}
  \caption{GAL encoding of a channel carrying messages (5--26) featuring a type (8) and an address field (7). Part of the GAL modeling the L1 cache (27--46).}
  \label{fig:changal}
\end{figure}

The processor is modeled as a three state automaton, alternating between an idle state and a state where a read (resp. write) request is sent to the L1 cache on an arbitrary address. Then, it awaits the reply to go back to idle. The Memory cells are even simpler: since data values have been abstracted away, they feature a single state and two transitions that simply acknowledge and reply appropriately to \texttt{PUT} and \texttt{GET} requests.

The L1 and L2 cache are much larger with respectively 14 and 16 states. They also feature variables that increase the state space size significantly. When transitions of an automaton read or write on channels, we add labels to those actions, indicating which channel is used. The message data is filtered by specifying values in the target label. For instance, the transition from \texttt{EMPTY} to \texttt{WRITE_WAIT_EMPTY} shown in Figure~\ref{fig:changal} reads a \texttt{DT_WR} request from the channel \texttt{PL1DTREQ} and writes a \texttt{WR} to the channel \texttt{L1L2DTREQ}. Some components such as the L1 cache have a set  identifier defined at initialization, that is used to tag outgoing messages, or to filter messages according to their target address. Transitions without communications on channels are left unlabeled, and can thus occur at any time.

\medskip
\textbf{Assembling a Configuration.}
From these models of channels and components we hierarchically build a representation of the full system.
A first composite type \texttt{CompositeCacheL1} is defined containing an instance of a Processor, an instance of a L1 cache, and two instances of channels connecting them together.

Channels are connected to appropriate endpoints using synchronizations. Unlabeled synchronizations are used to label local communications within the composite. The synchronization \texttt{s\_write\_PL1DTREQ} shown in Figure~\ref{fig:compo} is an example of this, and models the processor \texttt{p} sending a write request. The contents of the message (address and type) could be anything at this synchronization level. The labels that correspond to communications between the L1 cache and the L2 cache are however reexposed as labels of the composite, which simply forwards the request to the L1 cache instance.

We then build a top level composite acting as \code{main} that contains an array of instances of \texttt{CompositeCacheL1}, an array of instances of L2 caches and all the channel instances connecting them. The whole description is parametric and controlled by the three parameters set at the top of the file. Several other configurations were built to test each component in isolation.

\begin{figure}
\centering
	\begin{multicols}{2}
	\begin{lstlisting}
composite ProcessorCacheL1 {
 Processor p;
 CacheL1 c;
 ChannelAddrType chan_PL1DTREQ;
 ChannelAddrType chan_L1PDTACK;

 synchronization init (id_t $id)
  label "init" ($id) {
   c."init" ($id);
 }
 // local communications are unlabeled
 synchronization s_write_PL1DTREQ
  (addr_t $addr, type_t $type) {
   p."write_PL1DTREQ" ($addr, $type) ;
   chan_PL1DTREQ."write" ($addr, $type) ;
 }
 // exposing the ports of the L1 cache
 synchronization s_read_L2L1CPREQ
  (id_t $id, addr_t $addr, type_t $type)
  label "c_read_L2L1CPREQ"($id, $addr, $type) {
   c."read_L2L1CPREQ"($id,$addr, $type);
 }
...
\end{lstlisting}
 	\end{multicols}
  \caption{Composite encoding a component nesting a Processor, a L1 cache and two channels.}
  \label{fig:compo}
\end{figure}

\medskip
\textbf{Experiments and Measurements.}
On these models we were able to prove absence of deadlock and some logical properties expressed in CTL.
Overall we considered all configurations with up to \texttt{NB\_PROC} = 6 processors, up to \texttt{NB\_L2} = 3 L2 addresses, and a \texttt{CACHE\_TH} between 1 and 3. For each configuration, we ran the verification on an Intel Xeon 2.6 GHz machine with a limit set to 8 hours to complete the simulation, and a maximum of 192 GB of memory. Table~\ref{tab:results} reports on the experiments that finished within the time and memory constraints, and gives their time, the number of accessible states and the memory used.

The state space size is relatively modest, compared to the models crafted in other languages, thanks to the fine control over the atomicity of steps offered in GAL. We were indeed able to scale up to the configurations of interest, i.e. $3$ processors, $2$ L2 cache, and a threshold of $2$.

\begin{table}[ht!]
  \scriptsize
  \caption{Runtime (in seconds) and memory (in MB) required for the configurations of the parameters \texttt{NB\_PROC,NB\_L2,CACHE\_TH} which could be explored.}
  \label{tab:results}
  \begin{tabular}{|l|l|l|l|l|l|}
\hline
\texttt{PROC} & \texttt{L2} & \texttt{TH} &  States  & Time & Mem \\
\hline
1 & 1 & 1 &  51 & 0.05 & 5 \\
1 & 1 & 2 &  52 & 0.05 & 5 \\
1 & 1 & 3 &  53 & 0.06 & 5 \\
1 & 2 & 1 &  555 & 0.13 & 7 \\
1 & 2 & 2 &  565 & 0.12 & 7 \\
1 & 2 & 3 &  575 & 0.18 & 8 \\
1 & 3 & 1 &  4503 & 0.47 & 15 \\
1 & 3 & 2 &  4572 & 0.52 & 15 \\
1 & 3 & 3 &  4641 & 0.44 & 15 \\
2 & 1 & 1 &  7070 & 0.54 & 14 \\
2 & 1 & 2 &  1892 & 0.46 & 15 \\
2 & 1 & 3 &  2160 & 0.47 & 15 \\
2 & 2 & 1 &  681471 & 14 & 232 \\
2 & 2 & 2 &  68401 & 5 & 98 \\
2 & 2 & 3 &  77449 & 6 & 99 \\
2 & 3 & 1 &  2.76e+07 & 640.14 & 3482 \\
\hline
  \end{tabular}
  \begin{tabular}{|l|l|l|l|l|l|}
\hline
\texttt{PROC} & \texttt{L2} & \texttt{TH} &  States & Time & Mem  \\
\hline
2 & 3 & 2 &  1.13e+06 & 44 & 568 \\
2 & 3 & 3 &  1.27e+06 & 46 & 599 \\
3 & 1 & 1 &  175234 & 3 & 57 \\
3 & 1 & 2 &  226329 & 35 & 326 \\
3 & 1 & 3 &  130450 & 51 & 625 \\
3 & 2 & 1 &  4.32e+07 & 860.58 & 3871 \\
\textbf{3} & \textbf{2} & \textbf{2} &  6.77e+07 & 3692 & 7547 \\
3 & 2 & 3 &  1.54e+07 & 1396 & 6315 \\
3 & 3 & 3 &  5.14e+08 & 18759 & 66733 \\
4 & 1 & 1 &  3.36e+06 & 44 & 662 \\
4 & 1 & 2 &  5.53e+06 & 184 & 1645 \\
4 & 1 & 3 &  1.04e+07 & 574 & 3285 \\
4 & 2 & 2 &  4.49e+09 & 149471 & 171457 \\
5 & 1 & 1 &  6.06e+07 & 254 & 1389 \\
5 & 1 & 2 &  1.08e+08 & 9315 & 9439 \\
6 & 1 & 1 &  1.05e+09 & 831 & 3710 \\
6 & 1 & 2 &  1.87e+09 & 12324 & 23118 \\
\hline
  \end{tabular}
\end{table}

A set of $16$ properties were expressed in CTL, covering request response scenarios such as
"any time two processors share an address and one writes on it, the other one eventually gets an invalidate request."

\section{Conclusion}

We presented a case-study focusing on the modeling of a
cache-coherence protocol in GAL, and discussed previous
implementations using other languages and tools for the same
protocol. We noticed a certain difficulty, for the students who worked
on the subject, to assimilate the underlying formalism of each
tool. We also noticed how small changes in the language semantics can
result in big changes, either in the model description or in the state
space size. In particular, the communication and synchronisation primitives offered by a language are of high importance for getting clean and efficient models.

The parametric and compositional features of GAL proved adequate
to write models directly by hand. The parametric features are useful
both for writing a model with several instances of a same module
(e.g. cache) and for exploiting internally the similarities between
these modules to improve verification efficiency.

Overall, this case study showed that model-checking tools could
highlight real bugs and could run on problems of substantial size,
provided appropriate formal models can be built accurately.

\section*{Acknowledgements}
This work would not have been possible without the contributions of the students working on the project Mohamad Najem, Akli Mansour, Zahia Gharbi and Di Zhao. 

\bibliographystyle{eptcs}
\bibliography{mars2018}

\appendix
\newpage
\section{L1 Cache Finite State Machine}

\label{app:l1automaton}

\begin{figure}[ht!]
  \begin{center}
    \includegraphics[width=\columnwidth]{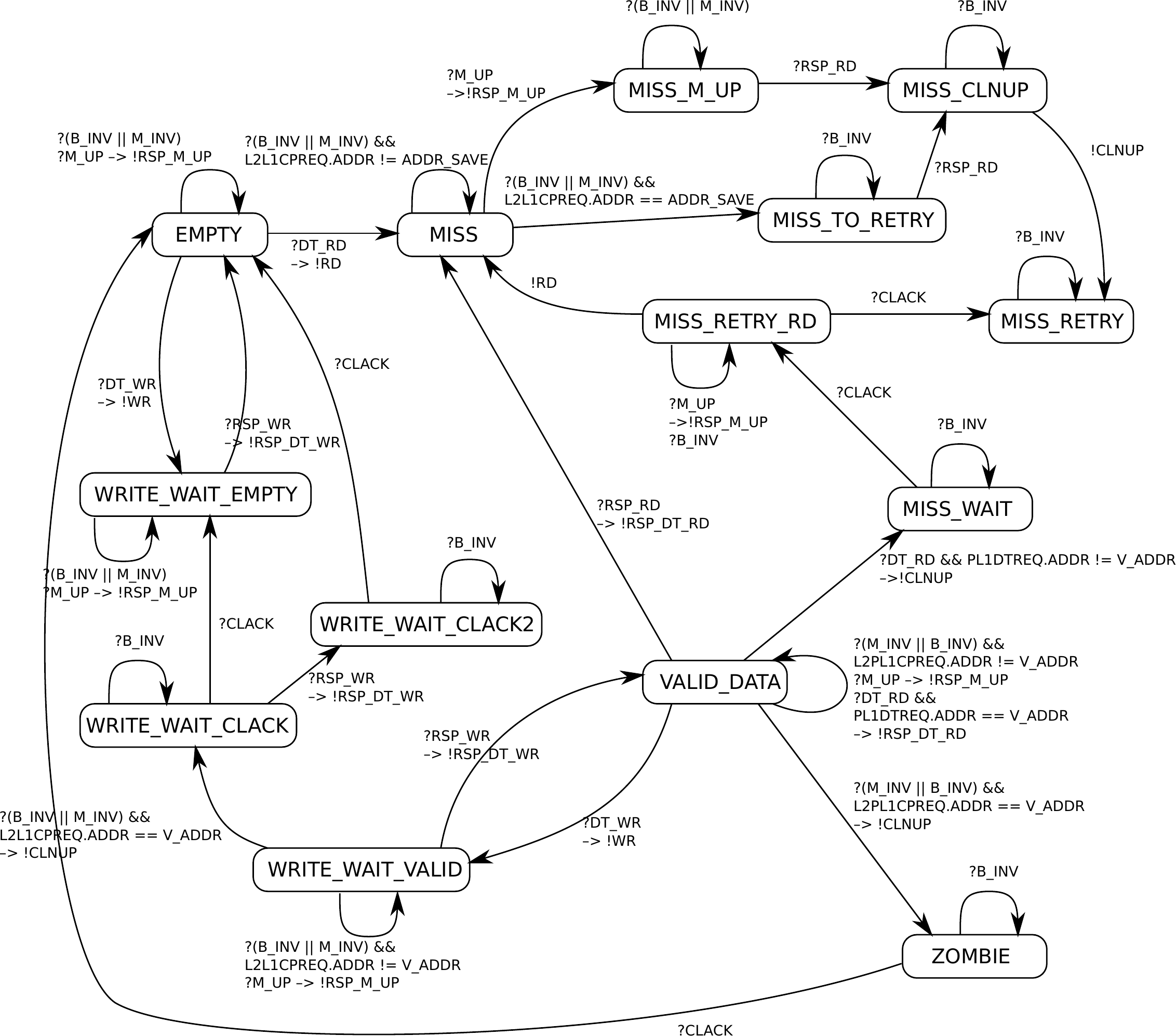}
  \end{center}
  \caption{L1 cache finite state machine. '?' denotes the reception of a message; '!' denotes the sending of a message; $\rightarrow$ denotes actions associated to a guard. Some actions are omitted for readability.}
  \label{fig:l1cache}
\end{figure}

\newpage
\section{L2 Cache Finite State Machine}
\label{app:l2automaton}

\begin{figure}[ht!]
  \begin{center}
    \includegraphics[width=\columnwidth]{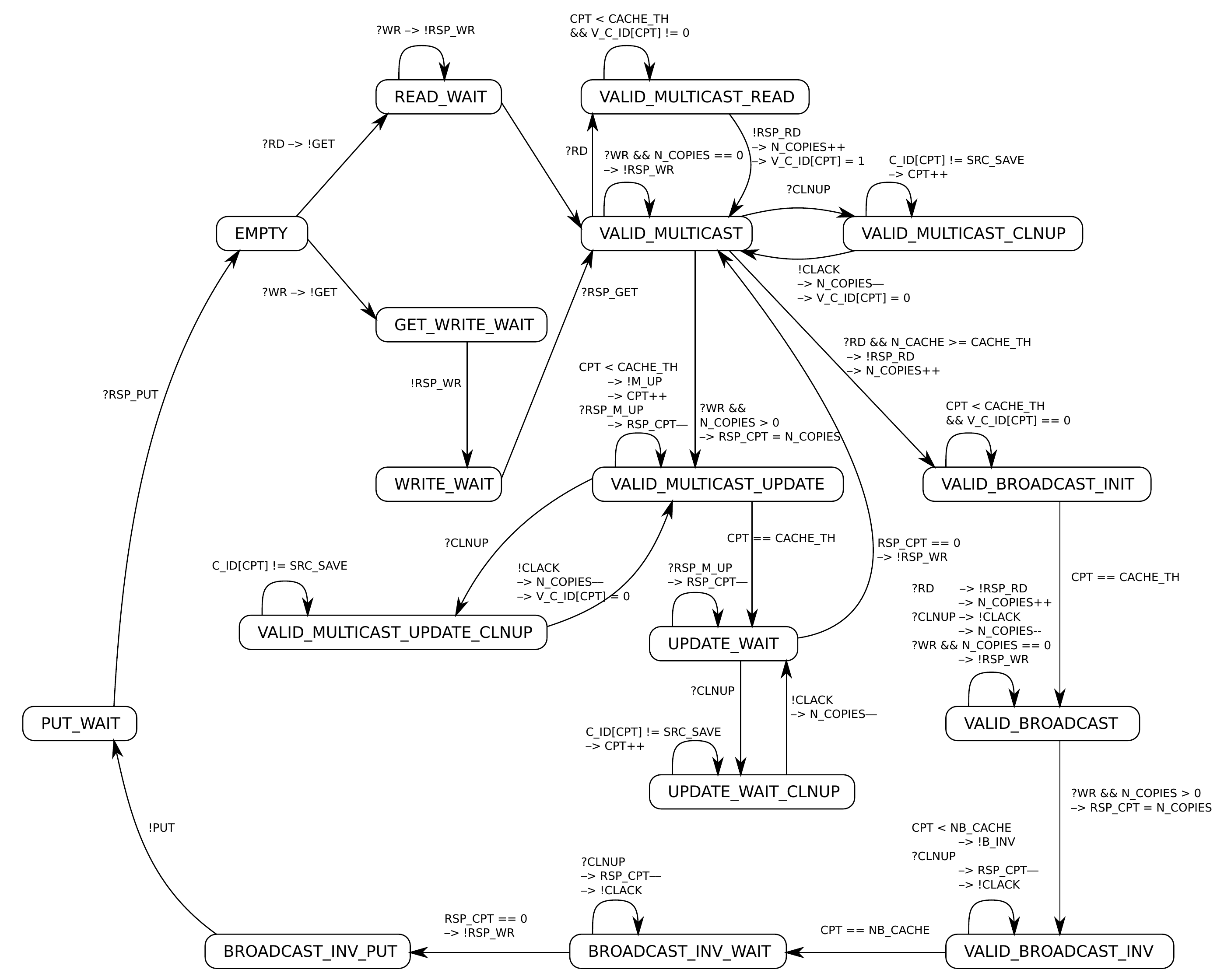}
  \end{center}
  \caption{L2 cache finite state machine. '?' denotes the reception of a message; '!' denotes the sending of a message; $\rightarrow$ denotes actions associated to a guard. Some actions are omitted for readability.}
  \label{fig:l2cache}
\end{figure}

\newpage
\section{Types of Messages and Associated Channels}
\label{app:messageschannels}

\begin{table}[ht!]
  \caption{Types of Messages and Channels on which they are Transported}
  \label{tab:messageschannels}
  \begin{center}
    \begin{tabular}{|l|l|l|}
      \hline
DT\_RD      & Read request from processor  & PL1DTREQ \\
DT\_WR      & Write request from processor & PL1DTREQ \\
RSP\_DT\_RD & Read response to processor   & L1PDTRSP \\
RSP\_DT\_WR & Write response to processor  & L1PDTRSP \\
RD          & Read request from L1 to L2   & L1L2DTREQ \\
WR          & Write request from L1 to L2  & L1L2DTREQ \\
RSP\_RD     & Read response from L2 to L1  & L2L1DTRSP \\
RSP\_WR     & Write response from L2 to L1 & L2L1DTRSP \\
CLNUP       & Line eviction from L1 to L2  & L1L2CPRSP \\
CLACK       & Line eviction received from L2 to L1 & L2L1CLACK \\
B\_INV      & Broadcast invalidate         & L2L1CPREQ \\
M\_INV      & Multicast invalidate         & L2L1CPREQ \\
M\_UP       & Multicast update             & L2L1CPREQ \\
RSP\_M\_UP  & Multicast update acknowledge & L1L2CPRSP \\
GET         & Read line from memory        & L2MEMDTREQ \\
PUT         & Write line to memory         & L2MEMDTREQ \\
RSP\_GET    & Response to GET request      & MEML2DTRSP \\
RSP\_PUT    & Response to PUT request      & MEML2DTRSP \\
      \hline
    \end{tabular}
  \end{center}
\end{table}

\end{document}